\documentclass[onecolumn,showpacs,nobibnotes,nofootinbib]{revtex4}
\usepackage{amsmath,amssymb}
\usepackage{graphicx}

\begin{document}

\title{Heavy-quark energy loss and thermalization in a strongly coupled SYM plasma}
\author{Guillaume Beuf}
\email{guillaume.beuf@cea.fr}
\affiliation{Institut de Physique Th{\'e}orique, CEA/Saclay, 91191 Gif-sur-Yvette cedex,
France \\
URA 2306, unit{\'e} de recherche associ{\'e}e au CNRS}
\author{Cyrille Marquet}
\email{cyrille@phys.columbia.edu}
\affiliation{Institut de Physique Th{\'e}orique, CEA/Saclay, 91191 Gif-sur-Yvette cedex,
France\\Department of Physics, Columbia University, New York, NY 10027, USA}
\author{Bo-Wen Xiao}
\email{bowen@phys.columbia.edu}
\affiliation{Department of Physics, Columbia University, New York, NY 10027, USA\\
Nuclear Science Division, Lawrence Berkeley National Laboratory, Berkeley, CA 94720, USA}

\begin{abstract}
Using the AdS/CFT correspondence, we compute the radiative energy loss of a slowly
decelerating heavy quark with mass $M$ moving through a supersymmetric Yang Mills (SYM)
plasma at temperature $T$ at large t'Hooft coupling $\lambda.$ The calculation
is carried out in terms of perturbation in $\sqrt{\lambda} T/M,$ and the rate
of the energy loss is computed up to second order. We explain the physical
meaning of each correction and estimate the thermalization time of a heavy
quark moving in a strongly-coupled plasma.
\end{abstract}


\maketitle


\section{Introduction}

Understanding the quark-gluon plasma created in nucleus-nucleus collisions
\cite{brahms,phobos,star,phenix}
at the Relativistic Heavy Ion Collider (RHIC) is a challenging task. Many measurements
have been carried out to provide insight on the properties of that dense QCD matter,
yet a few years after the discovery, the question whether the plasma formed at RHIC
is weakly or strongly coupled remains unanswered. On the one hand, results on bulk
quantities like the elliptic flow $v_2$ \cite{Molnar:2001ux} or the shear viscosity to entropy ratio $\eta/s$ \cite{Policastro:2001yc,Kovtun:2004de,Teaney:2003kp}
fit within a perfect-liquid picture. The large flow and low viscosity are naturally understood
from strong cross-sections. This quickly prompted claims that the plasma is strongly
coupled, although from a broader perspective, these issues are still being debated.
On the other hand, one would like to establish a global picture and be able to also
answer the question using hard probes.

Hard probes are believed to be ideal processes to study the properties of the quark-gluon
plasma, they are thought to be understood well enough to provide clean measurements.
Observables built to measure medium effects on particle production, like the nuclear
modification factors $R_{AA},$ are not easily reproduced in perturbative QCD (pQCD),
even for hard production. For instance to reproduce light-hadron $R_{AA}$'s within the
pQCD framework of medium-induced energy loss
\cite{Baier:1996sk,Baier:1998kq,Zakharov:1996fv,Wiedemann:2000za,Gyulassy:2000er}, the value of the so-called jet quenching parameter $\hat{q}$ is adjusted to $5-10\ \mbox{GeV}^2/\mbox{fm}$
\cite{Eskola:2004cr}. It is otherwise estimated to be smaller ($1-3\ \mbox{GeV}^2/\mbox{fm}$) for a weakly coupled plasma at RHIC temperatures. Results on heavy-quark energy loss are also puzzling, high$-p_\perp$ electrons from charm and bottom mesons decays seem to indicate a similar suppression for light, charm and bottom quarks. By contrast in pQCD, the heavier the quark the weaker the suppression. While our current understanding of the pQCD picture must be still be improved (for instance collisional energy loss should be included as well), it is unclear if a pQCD approach can describe the suppression of high$-p_\perp$ particles at RHIC. Recent review and extensive discussion can be found in
Refs.~\cite{Abreu:2007kv,CasalderreySolana:2007zz,Zajc:2007ey}.

This motivates to think about strongly coupled plasmas. Addressing the strong coupling dynamics in QCD is an outstanding problem, the available tools are quite limited. However for the $\mathcal{N}=4$ Super Yang Mills (SYM) theory, the AdS/CFT correspondence
\cite{Maldacena:1997re,Witten:1998qj,Gubser:1998bc} is a powerful approach. Essentially, in the large$-N_c$ limit, the correspondence maps the quantum dynamics of the gauge theory at strong coupling into classical gravity dynamics in the fifth dimension of a curved space-time. While the
SYM theory is quite different from QCD (it is highly supersymetric and conformal), let us assume that the SYM plasma approximates well the QCD plasma, just above $T_c,$ the temperature of the
phase transiton. In fact it has been argued that some SYM results hold for any gauge theory with a gravity dual, like the lower bound of the shear viscosity \cite{Policastro:2001yc,Kovtun:2004de}. If the plasma formed at RHIC really is strongly-coupled, this insight gained in AdS/CFT calculations like the one in this paper could help developing a qualitative understanding. 

Recently, there have been a lot of AdS/CFT studies on heavy quark propagating in the SYM plasma. The energy loss is computed in Refs.~\cite{Herzog:2006gh,Gubser:2006bz}, the momentum broadening is calculated in Refs.~\cite{Gubser:2006nz,CasalderreySolana:2006rq,CasalderreySolana:2007qw}, the jet quenching parameter is evaluated in Ref.~\cite{Liu:2006he,Armesto:2006zv,Argyres:2006yz} and the stress-tensor of the system is derived in
Ref.~\cite{Friess:2006fk,Gubser:2007nd,Yarom:2007ni,Chesler:2007sv}. Moreover, the saturation momentum and structure functions of SYM plasma are studied in
Ref.~\cite{Hatta:2007he,Hatta:2007cs,Albacete:2008ze,Levin:2008vj}
and jet fragmentation and energy correlation are investigated in
Ref.~\cite{Hatta:2008tx,Hatta:2008tn,Hofman:2008ar,Noronha:2007xe}. Finally, there have been a few interesting studies which compare the predictions of heavy quark energy loss and jet correlations in pQCD with the ones in AdS/CFT
\cite{Horowitz:2007su,Betz:2008wy,Dominguez:2008vd,Marquet:2008kr,Mueller:2008zt,Hatta:2008st}.

In this work we study the propagation of a very energetic heavy quark through the SYM plasma at strong 't Hooft coupling and at temperature $T.$ Our starting point is the trailing-string picture of Herzog et al \cite{Herzog:2006gh}, which allows to compute the energy loss of the heavy quark in a static infinite-extend plasma. The heavy quark, in the fundamental representation, lives on a flavor brane close to the boundary with a string attached to it, hanging down to the horizon. The classical dynamics of the string is given by the Nambu-Goto action and from the string shape the energy loss can be determined. In \cite{Herzog:2006gh, Gubser:2006bz} a constant electric field is imposed on the flavor brane to drag the heavy quark at a constant speed $v.$
The electric field feeds energy into the string and is adjusted to compensate the energy lost
into the plasma. This calculation gives the leading order contribution to the energy loss when the mass of the heavy quark is extremely large. However, it is not such a realistic model.

In this paper we study what happens if one turns off the electric field, meaning if one stops pulling on the heavy quark. Then the energy loss of the quark is not compensated anymore, and the string actually losses energy, the heavy quark decelerates. Since we are working in the limit of large mass $M,$ the deceleration is slow and $\dot{v}\sim \sqrt{\lambda} T^2/M$ can be treated as a perturbation.
The deceleration modifies the shape of the string which in turns modifies the rate of energy loss by a correction of order $\sqrt{\lambda} T/M.$ This adds a higher order correction to $\dot{v}$ and so on. This is the problem that we solve. The calculation is carried out up to second order in
$\frac{\sqrt{\lambda} T}{M}.$ We explain the physical meaning of each correction which adds insights to the process of the heavy quark slowing down in a strongly coupled plasma.
Finally, in this picture the trailing string slowly gets straighter to eventually become that of a static quark. Of course we do not except our picture to apply that far. What will happen is that the heavy quark will eventually thermalize and when that happens, the discussion should be modified. However our calculation allows to estimate the thermalization time.

The plan of the paper in as follows. In Section II, we recall the string equations of motion derived from the Nambu-Goto action. In Section III, we discuss the possible ways the string gains and loses energy, and define the medium-induced energy loss which we are interested in. In Section IV, we compute the first and second order corrections to the trailing string picture, when one stops pulling on the heavy quark and it slows down. Finally Section V is devoted to an estimate of the thermalization time. Section VI concludes.

\section{The equations of motion of the string}

We consider the large $N_c,$ small gauge coupling $g_{YM}$ limit
\begin{equation}
N_c\to\infty\ ,\hspace{0.5cm} g_{YM}\to0\ ,\hspace{0.5cm} \lambda\equiv g_{YM}^2 N_c\mbox{ finite,}
\end{equation}
where the 't Hooft coupling $\lambda$ controls the $\mathcal{N}=4$ SYM theory. The strong couping regime means $\lambda \gg 1,$ and then the equivalent string theory in $AdS_5 \times S_5$ space is weakly coupled and weakly curved:
\begin{equation}
g_{YM}\ll1\Leftrightarrow g_s\ll1\mbox{ and }\lambda\gg1\Leftrightarrow R\gg l_s\ ,
\end{equation}
where $g_s$ is the string coupling, $l_s$ is the string length and $R$ is the curvature radius of the $AdS_5$ and $S_5$ spaces. In this limit of small string coupling and large curvature radius, classical gravity is a good approximation of the string theory and we can neglect excited modes along the $S_5$ space. The metric corresponding to the SYM theory at finite temperature is the
$AdS$ black brane solution in 5 dimensions which can be written as
\begin{equation}
ds^{2}=R^{2}\left\{ \frac{du^{2}}{u^{2}f(u)}-u^{2}f(u)dt^{2}
+u^{2}\left( dx^{2}+dy^{2}+dz^{2}\right) \right\}\ ,\quad f(u)=1-\frac{u_h^4}{u^4}\ ,
\end{equation}%
where $u_h=\pi T$ is a black-hole horizon. The corresponding Hawking temperature $T$ is the temperature of the SYM theory. The coordinate in the fifth dimension $u=r/R^2$ has the dimension of momentum and the SYM theory lives formally on the boundary at $u=\infty.$

In order to have one flavor of quarks in the fundamental representation in the field theory side, one introduces a D7 brane in the string theory side, whose embedding covers the part of the
$AdS_5$ space with $u\geq u_m= 2\pi m/\sqrt{\lambda}$, $m$ being the lagrangian mass of the heavy quark\footnote{In fact, the relation between $m$ and $u_m$ is modified at finite temperature. See Eq.(3.3) of Ref.\cite{Herzog:2006gh}. However, this modification will bring only higher order terms in our large mass expansion compared to the terms we calculate.}.
The heavy quark and the color field it generates are dual to a string attached to the D7 brane at $u=u_m$ and hanging down to the horizon. Points on the string can be identified to quantum fluctuations in the heavy quark wave function with virtuality
$\sim u.$ Indeed, the quantum dynamics in the SYM theory is mapped onto classical dynamics in the 5th dimension. More precisely, the string dynamics is given by the Nambu-Goto action
\begin{equation}
S=-T_{0}\int d\tau d\sigma \mathcal{L}=-T_{0}\int d\tau d\sigma \sqrt{-\det g_{ab}}
\end{equation}%
where $\left( \tau ,\sigma \right) $ are the string world-sheet
coordinates and $T_{0}$ is the string tension. We define
$X^{\mu}\left( \tau ,\sigma \right) $ as a map from the string
world-sheet to the five dimensional space time, and introduce the
following notations for derivatives:
$\dot{X}^{\mu }=\partial_{\tau }X^{\mu }$ and $X^{\prime \mu }=\partial _{\sigma }X^{\mu}$.
The determinant of the induced metric $-\det g_{ab}=-g$ is given by
\begin{equation}
-\det g_{ab}=\left( \dot{X}^{\mu }X_{\mu }^{\prime }\right) ^{2}-
\left( \dot{X}^{\mu }\dot{X}_{\mu }\right) \left( X^{\prime \mu }X_{\mu }^{\prime}\right) .
\end{equation}%
It is useful to define the string canonical momentum densities as follows
\begin{eqnarray}
\pi _{\mu }^{\tau } &=&-T_{0}\frac{\partial \mathcal{L}}{\partial \dot{X}%
^{\mu }}=-T_{0}\frac{\left( \dot{X}^{\nu }X_{\nu }^{\prime }\right) X_{\mu
}^{\prime }-\left( X^{\prime \nu }X_{\nu }^{\prime }\right) \dot{X}_{\mu }}{%
\sqrt{-g}}, \\
\pi _{\mu }^{\sigma } &=&-T_{0}\frac{\partial \mathcal{L}}{\partial X_{\mu
}^{\prime }}=-T_{0}\frac{\left( \dot{X}^{\nu }X_{\nu }^{\prime }\right) \dot{%
X}_{\mu }-\left( \dot{X}^{\nu }\dot{X}_{\nu }\right) X_{\mu }^{\prime }}{%
\sqrt{-g}}.
\end{eqnarray}%
Now let us consider an arbitrary variation $\delta X^{\mu }$ in the action.
After neglecting the surface terms, we can obtain the equations of motion of
the string\ according to the variation principle%
\begin{equation}
\frac{\delta S}{\delta X^{\mu }}=0\Rightarrow \frac{\partial \mathcal{L}}{%
\partial X^{\mu }}-\frac{d}{d\tau }\frac{\partial \mathcal{L}}{\partial \dot{%
X}^{\mu }}-\frac{d}{d\sigma }\frac{\partial \mathcal{L}}{\partial X^{\prime\mu}}=0.  \label{elequation}
\end{equation}

When one chooses a static gauge by setting $\left( \tau ,\sigma \right)
=\left( t,u\right) $, and defines $X^{\mu }=\left( t,u,x\left( t,u\right)
,0,0\right) $ (we assume that the quark moves along the $x$ direction), it is straightforward to find that%
\begin{equation}
-\det g_{ab}=R^{4}\left( 1-\frac{\dot{x}^{2}}{f\left( u\right) }%
+u^{4}f\left( u\right) x^{\prime 2}\right) .
\end{equation}%
The string canonical momentum densities reduce to
\begin{equation}
\left(
\begin{array}{c}
\pi _{x}^{\tau } \\
\pi _{u}^{\tau } \\
\pi _{t}^{\tau }%
\end{array}%
\right) =\frac{T_{0}R^{4}}{\sqrt{-g}}\left[
\begin{array}{c}
\dot{x}\,\,f^{-1} \\
-\dot{x}\,x^{\prime }\,\,f^{-1} \\
-1-(x^{\prime })^{2}\,u^{4}\,f%
\end{array}%
\right] ,\quad \left(
\begin{array}{c}
\pi _{x}^{\sigma } \\
\pi _{u}^{\sigma } \\
\pi _{t}^{\sigma }%
\end{array}%
\right) =\frac{T_{0}R^{4}}{\sqrt{-g}}\left[
\begin{array}{c}
-x^{\prime }\,u^{4}\,f \\
-1+(\dot{x})^{2}\,f^{-1} \\
\dot{x}\,x^{\prime }\,u^{4}\,f%
\end{array}%
\right] .  \label{densities}
\end{equation}%
For $X^{\mu }=x\left( t,u\right)$ and $X^{\mu }=t,$ Eq.~(\ref{elequation}) gives
two equations which represent the conservation of momentum and energy, respectively:
\begin{eqnarray}
\frac{\partial\pi _{x}^{\tau }}{\partial\tau }+\frac{\partial\pi _{x}^{\sigma }}{\partial\sigma }&=&0,
\label{ex} \\
\frac{\partial\pi _{t}^{\tau }}{\partial\tau }+\frac{\partial\pi _{t}^{\sigma }}{\partial\sigma }&=&0.
\label{et}
\end{eqnarray}%
For $X^{\mu }=u$, one obtains
\begin{equation}
\frac{\partial\pi _{u}^{\tau }}{\partial\tau }+\frac{\partial\pi _{u}^{\sigma }}{\partial\sigma }=-T_{0}%
\frac{2}{\mathcal{L}}\left( \frac{\dot{x}^{2}}{f^{2}\left( u\right) }\frac{%
u_{h}^{4}}{u^{5}}+u^{3}x^{\prime 2}\right) .  \label{eu}
\end{equation}%
Due to the curvature in the fifth dimension, $\frac{\partial L}{\partial u}$
generates a tidal force on the string in the $u$ direction. However, among
those three equations of motion above, there is only one independent
equation due to the gauge fixing. One can just solve Eq. ~(\ref{ex}), and
check that the solution $x\left( t,u\right) $ is also the solution of the
other two. Therefore, derived from Eq.~(\ref{ex}), the equation of motion of the
classical string in the $x$ direction can be written as
\begin{equation}
\frac{\partial }{\partial u}\left( \frac{u^{4}f\left( u\right) x^{\prime }}{%
\sqrt{-g}}\right) -\frac{1}{f\left( u\right) }\frac{\partial }{\partial t}%
\left( \frac{\dot{x}}{\sqrt{-g}}\right) =0.  \label{eom}
\end{equation}


\section{The energy of the string during the heavy quark deceleration}

The energy flow along the string at the position $u=\sigma$ towards smaller $u$ is given by
\begin{equation}
\Phi(t,u)\equiv\pi _{t}^{\sigma}=\frac{%
T_{0}R^{4}}{\sqrt{-g}}\dot{x}x^{\prime }u^{4}f(u). \label{eflow}
\end{equation}%
The energy density of the string is given by $\pi _{t}^{\tau}$ as:
\begin{equation}
e(t,u)\equiv-\pi _{t}^{\tau}=\frac{%
T_{0}R^{4}}{\sqrt{-g}}\left[1+x^{\prime 2}u^{4}f(u)\right].  \label{edensity}
\end{equation}

We consider a string which moves at a velocity $v$ in $AdS_{5}.$ Generally speaking, at finite temperature and velocity,
there is always an event horizon in the fifth dimension which divides the
trailing string into two parts. The upper part genuinely belongs to the dual of the heavy quark and its wave function,
while the lower part is part of the plasma. This horizon is located at\footnote{%
This has been shown explicitly in Ref.~\cite{Dominguez:2008vd}. $u_s$ is
also the location where the local speed of light in AdS space coincides with
the propagation speed of the string. From the heavy quark point of view, the
event horizon is at $u_s$.} $u=u_s$ where
\begin{equation}
u_s=\sqrt{\gamma}u_h\mbox{ with }\gamma=\frac{1}{\sqrt{1-v^2}}\ .
\label{satscale}\end{equation}
On the gauge theory side, this means that the part of the string above $u_s$
corresponds to highly virtual fluctuations part of the heavy quark wave function
while the part of the string below $u_s$ corresponds to longer-lived
fluctuations which became emitted radiation. In the following we study the system formed by the upper part of the string, dual to the heavy quark dressed by the highly virtual fluctuations. Therefore, we consider the energy of upper part of the string
\begin{equation}
E(t)=\int_{u_s}^{u_m} du\ e(t,u)=-\int_{u_s}^{u_m} du\ \pi_t^\tau = T_0 R^4 \int_{u_s}^{u_m} du\ \frac{1+u^4
f(u) x^{\prime 2}}{\sqrt{-g}}\ .  \label{defE}
\end{equation}

Then, from \eqref{defE}, one calculates the energy change of the system
during $dt$
\begin{eqnarray}
dE &=&d\left( -\int_{u_{s}}^{u_{m}}du\ \pi _{t}^{\tau}(t,u)\right)  \notag \\
&=&\pi _{t}^{\tau }(t,u_{s})\ d(u_{s})-dt\int_{u_{s}}^{u_{m}}du\ \partial
_{t}\pi _{t}^{\tau }(t,u)  \notag \\
&=&\pi _{t}^{\tau }(t,u_{s})\ d(u_{s})+dt\int_{u_{s}}^{u_{m}}du\ \partial
_{u}\pi _{t}^{\sigma }(t,u)  \notag \\
&=&\pi _{t}^{\tau }(t,u_{s})\ d(u_{s})-\pi _{t}^{\sigma }(t,u_{s})dt+\pi
_{t}^{\sigma }(t,u_{m})dt\ ,
\end{eqnarray}%
where we have used Eq.~(\ref{et}) to go from the second to the third line.
Hence, the global energy conservation equation of the system during $dt$
reads
\begin{equation}
dE=\delta \left. E\right\vert _{du_{s}}-\Phi(t,u_s)\  dt +\Phi(t,u_m)\  dt,
\label{globalEBalance}
\end{equation}%
where
\begin{equation}
\delta \left. E\right\vert _{du_{s}}=\pi _{t}^{\tau }(t,u_{s})\ \frac{%
u_{s}\gamma ^{2}v\dot{v}}{2}dt=e(t,u_s)\  \frac{%
u_{s}\gamma ^{2}v(-\dot{v})}{2}dt \label{deltaEdus}
\end{equation}%
is the energy gain of the system during $dt$ due to the increase of the size of our system. Indeed, $u_{s}$ decrease as $v$ decrease.
\begin{equation}
\Phi(t,u_s)\  dt=\pi _{t}^{\sigma
}(t,u_{s})\ dt  \label{deltaEradus}
\end{equation}%
is the energy loss of the system during $dt$ due to the energy flow along
the string at $u=u_{s}$, and
\begin{equation}
\Phi(t,u_m)\  dt=\pi _{t}^{\sigma
}(t,u_{m})\ dt  \label{deltaEradum}
\end{equation}%
is the energy gain of the system during $dt$ due to the incoming energy flow at the top
of the string at $u=u_{m}$.

In the heavy quark energy loss problem, these contributions can be interpreted as follows:
$\Phi(t,u_m)\  dt$ is the work of external forces acting on the heavy quark and
$\Phi(t,u_s)\  dt$ is the energy radiated by the dressed heavy quark: fluctuations softer than
the scale given by $u_s$ are freed into the plasma. As the heavy quark decelerates, $u_s$ becomes softer and more fluctuations are kept in the wave function. This leads to the energy gain $\left.\delta E\right\vert _{du_{s}}.$

In Ref.~\cite{Herzog:2006gh}, an external force is acting on the heavy quark and
adjusted in order to have a stationary solution, with $\dot{v}=0$ and $dE/dt=0$ for the upper part of the string. Thus, the energy flow is uniform along the string, meaning
$\Phi(t,u_s)=\Phi(t,u_m).$
The same amount of energy is put into the string at the top than flows
at the bottom into the horizon, or equivalently is radiatively lost by the heavy quark.
However, this model does not describe the heavy quark energy loss so satisfactorily,
since in reality the heavy quark will slow down when propagating in the plasma.

In this paper we would like to consider the case where the external
force is turned off and the heavy quark is allowed to decelerate slowly. In
this case, we always keep $\Phi(t,u_m)$ vanishing, and we compute the energy loss. 
Although the total energy loss $-dE/dt$ of the upper part of the string includes the contribution
from the term $\delta \left. E\right\vert _{du_{s}},$ we focus on the radiative energy loss
$\Phi(t,u_s)$ which is physically more important. For instance, this contribution could be
compared to calculations performed in the pQCD framework
\cite{Baier:1996sk,Baier:1998kq,Zakharov:1996fv,Wiedemann:2000za,Gyulassy:2000er}.

\section{Perturbations of the trailing string due to deceleration}

\subsection{The stationary solution as the leading contribution in a large quark mass expansion.}

Let us review briefly the main result of Ref.~\cite{Herzog:2006gh,Gubser:2006bz}, which is the stationary solution corresponding to a the heavy quark ($u_m\to\infty$) pulled by a external force imposing a constant velocity. The stationarity condition gives the Ansatz
\begin{equation}
x=x_{0}+vt+vF(u)\text{ with }\dot{v}=0\ .\label{Ansatz_0thOrder}
\end{equation}
Since there is no time dependence in both $\dot{x}$ and $x^{\prime }$, the
time derivative gives vanishing result in the equation of motion. Thus, one
can equate both the space-dependent and time-dependent parts of Eq.~(\ref%
{eom}) to zero, respectively. Therefore, one obtains
\begin{equation}
x^{\prime 2}=\frac{C_{0}^{2}}{\left[ u^{4}f\left( u\right) \right] ^{2}}%
\frac{1-\frac{u_{h}^{4}}{u^{4}}-v^{2}}{1-\frac{u_{h}^{4}}{u^{4}}-\frac{%
C_{0}^{2}}{u^{4}}},
\end{equation}%
where $C_{0}$ is an integration constant. In order to maintain finite and
positive value of $x^{\prime 2}$, we should require that the numerator and
denominator change sign at the same $u$ value ($u=u_{s}=\sqrt{\gamma }u_{h}$).
Thus, this fixes the constant $C_{0}=u_{h}^{2}v\gamma $, and yields
$x^{\prime }=\pm \frac{u_{h}^{2}v}{u^{4}f\left( u\right) }$, where the $+$
sign corresponds to a flow into the horizon at $u_{h}$ and the $-$ sign
corresponds to a flow out of the horizon. By requiring the incoming boundary
condition, we pick up the solution with the $+$ sign and discard the other
solution.

Then, the shape of the string is found to be
\begin{equation}
F(u)=\frac{1}{2u_{h}}\left[ \frac{\pi }{2}-\tan ^{-1}\left( \frac{u}{u_{h}}%
\right) -\coth ^{-1}\left( \frac{u}{u_{h}}\right) \right] \quad \text{ with}%
\quad vF^{\prime }(u)=\frac{vu_{h}^{2}}{u^{4}-u_{h}^{4}}\label{sol_F}
\end{equation}%
and with the boundary condition $\left. F\left( u\right) \right\vert _{u=\infty}=0 $.
Using that solution, one can compute the energy density of the string and the energy flow
along the string:
\begin{eqnarray}
e(u)&=&\frac{T_0 R^4}{\sqrt{-g}}\ [1+u^4f(u) x^{\prime 2}]
=\frac{T_0 R^2}{\gamma}\frac{\gamma^2 u^4-u_h^4}{u^4-u_h^4}
\label{e_Stat}\ ,\\
\Phi(u)&=&\frac{T_0R^4}{\sqrt{-g}}\ \dot{x}x^{\prime }u^{4}f(u)
=T_0 R^2\ \gamma v^2 u_h^2
\label{Phi_Stat}\ .
\end{eqnarray}
The energy flow along the string is constant, and the energy stored in the upper part of the string ($u>u_s$) is
\begin{equation}
E= \int_{u_s}^{u_m} du\ e(u)=T_0 R^2\ \gamma
\left\{u_m-u_s +v^2 u_h^2 [F(u_m)-F(u_s)]\right\}\ .
\label{E_Stat}
\end{equation}
The energy conservation equation \eqref{globalEBalance} is trivially satisfied:
$dE/dt=\delta \left. E\right\vert_{du_s}/dt=0$ and $\Phi(u_m)=\Phi(u_s).$

When the external force is turned off, $\Phi(t,u_m)=0$ and in order to keep
\eqref{globalEBalance} satisfied, $\dot{v}$ must be non zero (unless the mass of the heavy
quark is stricly infinite). Indeed $dE/dt$ is no longer zero:
\begin{equation}
\frac{dE}{dt}=T_0R^2\gamma^3 v\dot{v}\left[u_m-u_s+\frac{\gamma^2+1}{\gamma^2}u_h^2(F(u_m)-F(u_s))-\frac{\gamma^2+1}{2\gamma^2}u_s\right]+{\cal O}(\dot{v}^2)\ ,
\label{dedt}\end{equation}
and this imposes that in the large $u_m$ limit, $-\dot{v}\propto u_h^2/u_m$
\footnote{The ${\cal O}(\dot{v}^2)$ terms come from the fact that after switching off
the external field, the energy of the string differs from \eqref{E_Stat} by a
contribution of order $\dot{v}.$}.
This follows from the fact that in the energy conservation equation \eqref{globalEBalance}, the finite term (when $u_m\to\infty$) $\Phi(t,u_s)=T_0 R^2\ \gamma v^2 u_h^2+{\cal O}(\dot{v})$ in the right-hand side must be compensated by a finite piece in $dE/dt.$ There is indeed no finite terms coming from the other contribution
\begin{equation}
\frac{\delta \left. E\right\vert _{du_{s}}}{dt}=-e(t,u_s)\
\frac{u_s\gamma ^{2}v\dot{v}}{2}=-T_0 R^2\ \ \frac{\gamma(\gamma^2+1)}{2}\ v\  \dot{v}\ u_s
+{\cal O}(\dot{v}^2)\ .\label{dus}
\end{equation}
Therefore we now treat $\dot{v}$ as a small parameter in the heavy quark deceleration problem.
On the field theory side, this corresponds to a small $(T\sqrt{\lambda})/(2 m)$ expansion according to the AdS/CFT dictionary.

\subsection{First order in solving the equation of motion}


Since the deceleration is very slow for heavy quarks, one can view the string as a quasi-static trailing string at every moment. Its shape should not be too different from the trailing string solution found above, except close to $u=u_m$ where corrections are of order one, as is the case for $\Phi(t,u_m)$ which is now vanishing. Thus, we perform a perturbative expansion of the shape
$x(t,u)$ of the string as a series in the small parameter $\dot{v}/u_h$. In our quasi-static picture, we can assume that the coefficients of the expansion depend on time only through $v(t)$. Hence, we write the generalization of the Ansatz \eqref{Ansatz_0thOrder} including the first order correction as
\begin{equation}
x(t,u)=x_{0}+\int^t v(t')dt'+v(t)F(u)+\dot{v}(t)\ \zeta (u,v(t))\ .\label{Ansatz_1stOrder}
\end{equation}
After neglecting the $\dot{v}^{2}$ and $\ddot{v}$ terms, which are
suppressed by one more factor of $\frac{u_{h}}{u_{m}}$, we reach the
equation of motion at first order in $\frac{u_{h}}{u_{m}}$ (the terms which
are proportional to $\dot{v}$).
\begin{equation}
\gamma \dot{v}\frac{\partial }{\partial u}\left[ \left( u^{4}f(u)-\gamma
^{2}v^{2}u_{h}^{4}\right) \zeta ^{\prime }(u,v)+\gamma ^{2}v^{2}u_{h}^{2}%
\frac{F(u)}{f(u)}\right] =\frac{\gamma ^{3}\dot{v}}{f(u)}.  \label{eom1}
\end{equation}%
Solving Eq.~(\ref{eom1}) yields
\begin{equation}
\zeta ^{\prime }(u,v)=\frac{\gamma ^{2}\left[ u+u_{h}^{2}F(u)-v^{2}u_{h}^{2}%
\frac{F(u)}{f(u)}-C_1\right] }{u^{4}-\gamma ^{2}u_{h}^{4}}= \frac{\gamma ^{2}\left[ u-C_{1}\right] }{u^{4}-u_{s}^{4}}+F^{\prime }(u) \ F(u)\ .\label{zetaprime}
\end{equation}
As for the previous order, we determine the integration constant $C_{1}$ by requiring a regular shape of the string at $u=u_s$, and thus $C_{1}=u_s$.
After the final integration and requiring that $\left. \zeta (u)\right\vert
_{u=\infty }=0$, one gets
\begin{equation}
\zeta (u,v)=\frac{\gamma ^{2}}{2u_{s}^{2}}\left[ \tan ^{-1}\left( \frac{u}{%
u_{s}}\right) +\log \left( \frac{u+u_{s}}{\sqrt{u^{2}+u_{s}^{2}}}\right) -%
\frac{\pi }{2}\right] +\frac{F^{2}\left( u\right) }{2}\ .\label{sol_zeta}
\end{equation}
Strictly speaking the conditions $\left. \zeta (u)\right\vert
_{u=\infty }=0$ and $\left. F (u)\right\vert
_{u=\infty }=0$ imply that $v(t)$ is not really the velocity of the quark, which is supposed to live at $u=u_m$. However, as we show \emph{a posteriori} in appendix, the difference is relevant only at higher orders in our large mass expansion. 

For any arbitrary point on the string, one can compute the energy flow $\Phi(t,u)$ which
runs through it from the top part of the string to the bottom. First at $u=u_m,$ all the contributions in $x'$ are of the same order, and one obtains
\begin{eqnarray}
\Phi(t,u_m)&=&\left. -\frac{T_{0}R^{4}}{\sqrt{-g}}
\dot{x}x^{\prime }u^{4}f(u)\right\vert _{u=u_{m}} \\
&=&-\frac{T_{0}R^{2}}{\sqrt{1-v^{2}}}v\left\{ vu_{h}^{2}
+\dot{v}\zeta^{\prime }(u_m,v)u_m^{4}f\left(u_m\right) \right\}\ .
\end{eqnarray}%

Imposing the Neumann boundary condition $\Phi(t,u_m)=0$ on the flavor brane gives
\begin{equation}
0=T_{0}R^{2}\gamma v^2 u_h^2 \left[1+ \frac{\dot{v}\gamma^2 (u_m-u_s)}{v u_h^2}+ {\cal O}\left(\frac{\dot{v}^2}{u_h^2}\right) \right]\ .\label{Phi_um_1stOrder}
\end{equation}
The deceleration of the string at first order is then given by
\begin{equation}
\dot{v}(t)=-\frac{v\ u_h^2}{\gamma^2 (u_m-u_s)}\ .\label{vdot_1stOrder}
\end{equation}
This equation allows us to specify our expansion parameter, it is actually $u_h/(u_m-u_s),$ and not $u_h/u_m.$ In particular, our perturbative expansion breaks down if $u_s$ is too close to
$u_m,$ meaning if the point which determines the energy loss is too close to the flavor brane\footnote{For $u\lesssim u_m,$ this string picture is modified anyway
\cite{Callan:1997kz,Gibbons:1997xz,Chesler:2007sv}. The string is actually a narrow tube: the interaction of the string with the D7 brane deforms the brane in the neighborhood of the endpoint of the string over a length scale of order $1/M.$}. In order to interpret this parameter on the gauge theory side, let us come back to the energy of the upper part of the string. The first
order contribution is
\begin{equation}
E(t)=T_0 R^2\ \gamma (u_m-u_s)\left[1+{\cal O}\left(\frac{u_h}{u_m-u_s}\right)\right]\ .
\end{equation}
On the gauge theory side, this is the energy of the color field of the quark. It is smaller than in the vacuum due to the screening of soft fluctuations by the plasma. As in
Ref.~\cite{Herzog:2006gh,Chernicoff:2008sa}, we shall use an abuse of language and denote
$E/\gamma$ the {\it thermal} mass of the quark $M.$ It is different from the Lagrangian mass $m$ because of the finite temperature, but the interpretation as a quark mass on the field theory
side is not entirely clear. In particular, while for the heavy quark at rest all three definitions of the mass $M$ coincide, for a moving quark the $\gamma$ dependence proposed in
\cite{Herzog:2006gh,Chernicoff:2008sa} is different from ours. In our case, the $\gamma$ dependence accounts for the fact that as the heavy quark decelerates, softer modes are allowed (not screened by the plasma anymore) in his wavefunction:
\begin{equation}
M=\frac{\sqrt{\lambda}}{2\pi}(u_m-u_s)\ .
\end{equation}
It is also defined so that $-Md\gamma/dt=\Phi(t,u_s),$ the radiative energy loss, which for now can be checked at leading order. In terms of $M,$ we can now write our expansion parameter as $(T\sqrt{\lambda})/(2 M).$ In appendix, we give the second order and third order (this one is obtained from the correction $\zeta(u,v)$ to the trailing string) corrections to the energy and therefore to the mass.

Now we can compute the energy flow for $u\ll u_{m},$ using equations
\eqref{Ansatz_1stOrder} and \eqref{zetaprime}:
\begin{equation}
\Phi(t,u)=\frac{T_{0}R^{4}}{\sqrt{-g}}\dot{x}x^{\prime }u^{4}f(u)=T_{0}R^{2}\gamma v^2 u_h^2 \left[1+(1+\gamma^2) \frac{\dot{v}}{v} F(u)+ \frac{\dot{v}\gamma^2 (u-u_s)}{v u_h^2}+ {\cal O}\left(\frac{\dot{v}^2}{u_h^2}\right) \right]
. \label{Phi_1stOrder}
\end{equation}
By evaluating equation \eqref{Phi_1stOrder} at $u=u_s$ and inserting equation \eqref{vdot_1stOrder}, one finds the outgoing energy flow from the upper part of the string
\begin{equation}
\Phi(t,u_s)=T_{0}R^{2}\gamma v^2 u_h^2 \left[1-\frac{(1+\gamma^2) u_h^2 F(u_s)}{\gamma^2 (u_m-u_s)}+ {\cal O}\left(\frac{u_h^2}{(u_m-u_s)^2}\right) \right]. \label{Phi_us_1stOrder}
\end{equation}
Using formulae \eqref{dus} and \eqref{Phi_us_1stOrder}, it is easy to see that the energy conservation equation
\begin{equation}
\frac{dE}{dt}=-\Phi(t,u_s)+\frac{\delta \left. E\right\vert _{du_{s}}}{dt}
\end{equation}
is verified, and to recover the net energy loss of the system at first order \eqref{dedt}.
Using the AdS/CFT dictionnary, one deduces the energy radiated as soft gluons by a dressed heavy quark moving through the SYM plasma at a speed $v$
\begin{eqnarray}
\Phi(t,u_s)&=&\frac{\sqrt{\lambda}}{2 \pi}\gamma v^2 (\pi T)^2
\left[1- \left(\frac{\sqrt{\lambda} T}{2M}\right)\frac{(1+\gamma^2) \pi T}{\gamma^2}
F(\sqrt{\gamma}\pi T) + {\cal O}\left(\frac{\lambda T^2}{4M^2}\right) \right]\\
&\underbrace{\simeq}_{\gamma\gg1}&\frac{\sqrt{\lambda}}{2 \pi}\gamma v^2 (\pi T)^2 \left[1+\frac1{3\gamma^{3/2}}\left(\frac{\sqrt{\lambda} T}{2M}\right)
+ {\cal O}\left(\frac{\lambda T^2}{4M^2}\right) \right]\ .\label{Phi_us_1stOrder_SYM}
\end{eqnarray}


As $F(u)$ is negative, the first order correction to the soft gluon emission $\Phi(t,u_s)$ is always positive, meaning that the rate of energy loss is stronger for lighter quarks.
Interestingly enough, in the ultrarelativistic case $\gamma\gg1$, the first order correction to $\Phi(t,u_s)$ is further suppressed by a factor $\gamma^{-3/2},$ which is unexpected. 
In the non-relativistic case $\gamma\gtrsim1$, due to the expansion
$F(u)\sim\frac{1}{4u_h}\log((u-u_h)/u_h)$ when $u\rightarrow u_{h}$, the first correction to 
$\Phi(t,u_s)$ is dominated by the logarithmically enhanced term. It shows that our large mass perturbative expansion breaks down at very small velocities. And even if due to the $v^2$ prefactor that logarithmic singularity does not lead to a divergent energy loss in the small velocity limit, but to a vanishing one as expected, in reality $\gamma$ never reaches 1. The heavy quark will thermalize before this happens and already then the picture has to be modified.



\subsection{Second order in solving the equation of motion}

At second order in $u_h/(u_m-u_s)$, we assume the solution of the
string to be of the following form
\begin{equation}
x(t,u)=x_{0}+\int^t v(t')dt'+v(t)F(u)+\dot{v}(t)\zeta (u,v(t))
+\dot{v}^2(t)\chi_1\left(u,v(t)\right) +\ddot{v}(t)\chi_2\left(u,v(t)\right)\ .
\label{Ansatz}
\end{equation}%
Thus one finds
\begin{eqnarray}
\dot{x}^{2} &=&v^{2}+2\dot{v}vF(u)+2\ddot{v}v\zeta (u,v)+\dot{v}%
^{2}F^{2}(u)+2v\dot{v}^{2}\frac{\partial \zeta (u,v)}{\partial v}+\cdots , \\
x^{\prime 2} &=&v^{2}F^{\prime 2}(u)+2\dot{v}vF^{\prime }(u)\zeta ^{\prime
}(u,v)+2v\dot{v}^{2}F^{\prime }(u)\chi _{1}^{\prime }\left( u,v\right) +2%
\ddot{v}vF^{\prime }(u)\chi _{2}^{\prime }(u,v)+\dot{v}^{2}\zeta ^{\prime
2}(u,v)+\cdots .
\end{eqnarray}
Expanding the time derivative part of the equation of motion up to second
order, one finds
\begin{eqnarray}
\frac{1}{f\left( u\right) }\frac{\partial }{\partial t}\left( \frac{\dot{x}}{%
\sqrt{-g}}\right)  &=&\frac{1}{f\left( u\right) }\left\{ \gamma ^{3}\dot{v}%
+\gamma \ddot{v}\left[ F(u)+\gamma ^{2}v^{2}\frac{F(u)}{f\left( u\right) }%
-\gamma ^{2}v^{2}u_{h}^{2}\zeta ^{\prime }(u,v)\right] \right\}   \notag \\
&&+\dot{v}^{2}\frac{v\gamma ^{3}F(u)}{f\left( u\right) }\left[ \frac{3\gamma
^{2}v^{2}}{f\left( u\right) }+\frac{2}{f\left( u\right) }+1\right] -\dot{v}%
^{2}\frac{v\gamma ^{3}u_{h}^{2}}{f\left( u\right) }\left[ \left( 2+3\gamma
^{2}v^{2}\right) \zeta ^{\prime }(u,v)+v\frac{\partial \zeta ^{\prime }(u,v)%
}{\partial v}\right]\ .
\end{eqnarray}%
Expanding the spatial derivative part yields
\begin{eqnarray}
\frac{\partial }{\partial u}\left( \frac{u^{4}f\left( u\right) x^{\prime }}{%
\sqrt{-g}}\right)  &=&\gamma \dot{v}\partial _{u}\left[ u^{4}f(u)\zeta
^{\prime }(u,v)-\gamma ^{2}v^{2}u_{h}^{4}\zeta ^{\prime }(u,v)+\gamma
^{2}v^{2}u_{h}^{2}\frac{F(u)}{f\left( u\right) }\right]   \notag \\
&&+\gamma \ddot{v}\partial _{u}\left[ u^{4}f(u)\chi _{2}^{\prime
}(u,v)-\gamma ^{2}v^{2}u_{h}^{4}\chi _{2}^{\prime }(u,v)+\gamma
^{2}v^{2}u_{h}^{2}\frac{\zeta (u,v)}{f\left( u\right) }\right]   \notag \\
&&+\gamma \dot{v}^{2}\partial _{u}\left[ u^{4}f(u)\chi _{1}^{\prime
}(u,v)-\gamma ^{2}v^{2}u_{h}^{4}\chi _{1}^{\prime }(u,v)+\gamma
^{2}vu^{4}f(u)\zeta ^{\prime }(u,v)\left( \frac{F(u)}{f\left( u\right) }%
-u_{h}^{2}\zeta ^{\prime }(u,v)\right) \right]   \notag \\
&&+\gamma \dot{v}^{2}\partial _{u}\left[ \gamma ^{2}vu_{h}^{2}\left( \frac{1%
}{2}\frac{F^{2}(u)}{f\left( u\right) }-\frac{1}{2}u^{4}f(u)\zeta ^{\prime
2}(u,v)+\frac{v}{f\left( u\right) }\frac{\partial \zeta (u,v)}{\partial v}%
\right) \right]   \notag \\
&&+\gamma \dot{v}^{2}\partial _{u}\left[ \frac{3}{2}\gamma
^{4}v^{3}u_{h}^{2}\left( \frac{F(u)}{f\left( u\right) }-u_{h}^{2}\zeta
^{\prime }(u,v)\right) ^{2}\right] .
\end{eqnarray}%
Comparing the coefficients of $\dot{v}^{2}$ and $\ddot{v}$ gives the
equations of motion for $\chi_1$ and $\chi_2$, respectively\footnote{
Strictly speaking, there is physically only one unknown function in our Ansatz
\eqref{Ansatz} at second order, $\dot{v}\chi_1+\ddot{v}\chi_2,$ but we choose to
determine $\chi_1$ and $\chi_2$ separately.}:%
\begin{eqnarray}
&&\partial _{u}\left[ u^{4}f(u)\chi _{1}^{\prime }(u,v)-\gamma
^{2}v^{2}u_{h}^{4}\chi _{1}^{\prime }(u,v)+\gamma ^{2}vu^{4}f(u)\zeta
^{\prime }(u,v)\left( \frac{F(u)}{f\left( u\right) }-u_{h}^{2}\zeta ^{\prime
}(u,v)\right) \right]   \notag \\
&&+\partial _{u}\left[ \gamma ^{2}vu_{h}^{2}\left( \frac{1}{2}\frac{F^{2}(u)%
}{f\left( u\right) }-\frac{1}{2}u^{4}f(u)\zeta ^{\prime 2}(u,v)+\frac{v}{f(u)%
}\frac{\partial \zeta (u,v)}{\partial v}\right) +\frac{3}{2}\gamma
^{4}v^{3}u_{h}^{2}\left( \frac{F(u)}{f\left( u\right) }-u_{h}^{2}\zeta
^{\prime }(u,v)\right) ^{2}\right]   \notag \\
&=&\frac{v\gamma ^{2}F(u)}{f\left( u\right) }\left[ \frac{3\gamma ^{2}v^{2}}{%
f\left( u\right) }+\frac{2}{f\left( u\right) }+1\right] -\frac{v\gamma
^{2}u_{h}^{2}}{f\left( u\right) }\left[ \left( 2+3\gamma ^{2}v^{2}\right)
\zeta ^{\prime }(u,v)+v\frac{\partial \zeta ^{\prime }(u,v)}{\partial v}%
\right] ,
\end{eqnarray}%
and
\begin{equation}
\partial _{u}\left[ u^{4}f(u)\chi _{2}^{\prime }(u,v)-\gamma
^{2}v^{2}u_{h}^{4}\chi _{2}^{\prime }(u,v)+\gamma ^{2}v^{2}u_{h}^{2}\frac{%
\zeta (u)}{f\left( u\right) }\right] =\frac{1}{f\left( u\right) }\left[
F(u)+\gamma ^{2}v^{2}\frac{F(u)}{f\left( u\right) }-\gamma
^{2}v^{2}u_{h}^{2}\zeta ^{\prime }(u,v)\right] .
\end{equation}%

First, let us integrate over $u$ once on both sides of the equations.
For $\chi_1^\prime(u,v)$ and $\chi_2^\prime(u,v)$, one gets, respectively:%
\begin{eqnarray}
\left( u^{4}-u_{s}^{4}\right) \chi _{1}^{\prime }(u,v) &=&-\gamma
^{2}vu^{4}f(u)\zeta ^{\prime }(u,v)\left( \frac{F(u)}{f\left( u\right) }%
-u_{h}^{2}\zeta ^{\prime }(u,v)\right)   \notag \\
&&-\gamma ^{2}vu_{h}^{2}\left( \frac{1}{2}\frac{F^{2}(u)}{f\left( u\right) }-%
\frac{1}{2}u^{4}f(u)\zeta ^{\prime 2}(u,v)+\frac{v}{f(u)}\frac{\partial
\zeta (u,v)}{\partial v}\right)   \notag \\
&&-\frac{3}{2}\gamma ^{4}v^{3}u_{h}^{2}\left( \frac{F(u)}{f\left( u\right) }%
-u_{h}^{2}\zeta ^{\prime }(u,v)\right) ^{2}-v\gamma ^{2}u_{h}^{2}\left[
2+3\gamma ^{2}v^{2}\right] \left[ \zeta (u,v)-\frac{1}{2}F^{2}\left(
u\right) \right]   \notag \\
&&-\gamma ^{4}vu_{h}^{2}\left[ 2+3\gamma ^{2}v^{2}\right] \varrho \left(
u,v\right) +3v\gamma ^{4}\sigma \left( u\right) -v^{2}\gamma ^{2}\phi \left(
u,v\right) +C_{2}\ ,\label{chi1eq}
\end{eqnarray}%
and
\begin{eqnarray}
\left(u^{4}-u_{s}^{4}\right) \chi _{2}^{\prime }(u,v) &=&-\gamma
^{2}v^{2}u_{h}^{2}\frac{\zeta (u,v)}{f\left( u\right) }+\gamma ^{2}\sigma
\left( u\right)  \\
&&-v^{2}u_{h}^{2}\gamma ^{2}\left[ \zeta (u,v)-\frac{1}{2}F^{2}\left( u\right) %
\right] -\gamma ^{4}v^{2}u_{h}^{2}\varrho \left( u,v\right) +C_{3}\ ,\label{chi2eq}
\end{eqnarray}%
where where $C_{2}$ and $C_{3}$ are integration constants. We have also introduced
the following functions
\begin{eqnarray}
\varrho \left( u,v\right)&=&u_{h}^{4}\int du
\frac{1}{u^{4}-u_{h}^{4}}\frac{u-u_{s}}{u^{4}-u_{s}^{4}}\\
\sigma \left( u\right)&=&\int du\frac{F\left( u\right) }{f\left(u\right)}
=\frac{u_{h}^{2}}{2}F^{2}\left( u\right) +uF\left( u\right)
+\frac{1}{4}\ln \frac{u^{2}+u_{h}^{2}}{u^{2}-u_{h}^{2}}\\
\phi \left( u,v\right)&=&u_{h}^{2}\int du\frac{1}{f\left( u\right) }\frac{%
\partial \zeta ^{\prime }(u,v)}{\partial v}
=\gamma u_{h}^{2}\int du\frac{1}{f\left( u\right) }
\frac{4u^{3}+3u^{2}u_{s}+2u_{s}^{2}u+u_{s}^{3}}{2\left( u+u_{s}\right) ^{2}
\left(u^{2}+u_{s}^{2}\right) ^{2}}\ ,
\end{eqnarray}
with
\begin{eqnarray}
\varrho \left( u,v\right)  &=&\frac{u_{h}}{4u_{s}^{2}\left(
u_{h}^{4}-u_{s}^{4}\right) }\left[ 2u_{s}^{3}\tan ^{-1}\left( \frac{u}{u_{h}}%
\right) -2u_{h}^{3}\tan ^{-1}\left( \frac{u}{u_{s}}\right) +u_{s}^{2}\left(
u_{h}-u_{s}\right) \log \left( u-u_{h}\right) \right]  \\
&&+\frac{u_{h}}{4u_{s}^{2}\left( u_{h}^{4}-u_{s}^{4}\right) }\left[
u_{s}^{2}\left( u_{h}+u_{s}\right) \log \left( u+u_{h}\right)
-2u_{h}^{3}\log \left( u+u_{s}\right) \right]  \\
&&+\frac{u_{h}}{4u_{s}^{2}\left( u_{h}^{4}-u_{s}^{4}\right) }\left[
-u_{s}^{2}u_{h}\log \left( u^{2}+u_{h}^{2}\right) +u_{h}^{3}\log \left(
u^{2}+u_{s}^{2}\right) \right] .
\end{eqnarray}
By requiring that $\chi _{1}^{\prime }(u)$ and $\chi _{2}^{\prime }(u)$ are
finite at $u=u_{s}$, one can determine the integration constants. They are such that in the
large $u$ limit for the right-hand side of \eqref{chi1eq} and \eqref{chi2eq}, one finds
\begin{eqnarray}
\left( u^{4}-u_{s}^{4}\right) \chi _{1}^{\prime }(u,v) &=&C_{2}^{\prime
}+\left( 5-\frac{5}{3\gamma ^{2}}\right) \frac{vu_{h}^{2}\gamma ^{6}}{u^{2}}+%
\mathcal{O}\left( \frac{u_{h}^{4}}{u^{4}}\right) +\cdots , \\
\left( u^{4}-u_{s}^{4}\right) \chi _{2}^{\prime }(u,v) &=&C_{3}^{\prime
}+\left( 1-\frac{5}{6\gamma ^{2}}\right) \frac{u_{h}^{2}\gamma ^{4}}{u^{2}}+%
\mathcal{O}\left( \frac{u_{h}^{4}}{u^{4}}\right) +\cdots ,
\end{eqnarray}%
with
\begin{eqnarray}
C_{2}^{\prime } &=&-\left( \frac{1}{4}+\frac{5\pi }{8}-\frac{5\log 2}{4}%
\right) v\gamma ^{5}, \\
C_{3}^{\prime } &=&-\left( \frac{\pi }{4}-\frac{\log 2}{2}\right) \gamma ^{3}\ .
\end{eqnarray}%

We shall not integrate the equations of motion any further to get the shape of the string, 
the knowledge of $\chi _{1}^{\prime }(u)$ and $\chi _{2}^{\prime }(u)$ is enough to compute
the energy flow $\Phi(t,u).$ First at $u=u_m,$ all the contributions in $x'$ are of the same order, and one obtains
\begin{eqnarray}
\Phi(t,u_m)&=&\left. \frac{T_{0}R^{4}}{\sqrt{-g}}
\dot{x}x^{\prime }u^{4}f(u)\right\vert _{u=u_{m}} \\
&=&-\frac{T_{0}R^{2}}{\sqrt{1-v^{2}}}v\left\{ vu_{h}^{2}+\left[ \dot{v}\zeta
^{\prime }(u_m,v)+\dot{v}^{2}\chi _{1}^{\prime }\left(u_m,v\right) +\ddot{v}%
\chi _{2}^{\prime }\left( u_m,v\right) \right] u_m^{4}f\left( u_m\right) \right\}\ .
\end{eqnarray}%
Imposing a vanishing flow at the top of the string gives the second order correction to $\dot{v}:$
\begin{equation}
\dot{v}=-\frac{vu_{h}^{2}}{\gamma^{2}(u_{m}-u_{s})}
+\frac{vu_{h}^{4}}{\gamma (u_{m}-u_{s})^{3}}
\left( \frac{1}{4}+\frac{\pi }{8}-\frac{\log 2}{4}\right)\ .\label{vdot2}
\end{equation}
Note that in the second order correction, we only gave the large $\gamma$ result\footnote{
While formula \eqref{satscale} is valid for the trailing string solution, the first order correction to the string shape modifies the $\gamma$ dependence of $u_s,$ the point where
the local speed of light in AdS space coincides with the propagation speed of the string. In principle this should be accounted for in the second order calculation, however in the large
$\gamma$ limit, these corrections are always subdominant.}, in order to get a lighter expression.
We will also need the second order expressions
\begin{equation}
\dot{v}^2=\frac{v^2(1-v^2)u_h^4}{\gamma^2(u_m-u_s)^2}\ ,\quad
\ddot{v}=\frac{v(1\!-\!3v^2)u_h^4}{\gamma^2(u_m-u_s)^2}\ .
\end{equation}

Now we can compute the energy flow for $u\ll u_{m}$ up to second
order in $u_h/(u_m-u_s),$ by expanding $\sqrt{-g}$, $\dot{x}$, and
$x^{\prime }u^{4}f(u)$ up to second order:
\begin{eqnarray}
\Phi(t,u)&=&\frac{T_{0}R^{4}}{\sqrt{-g}}\dot{x}%
x^{\prime }u^{4}f(u) \\
&=&T_0R^2\left( \pi T\right) ^{2}\frac{v^{2}}{\sqrt{%
1-v^{2}}}\left\{ 1+\left[ \frac{\gamma ^{2}\ u^{4}-u_{h}^{4}}{u^{4}-u_{h}^{4}%
}\right] \frac{\dot{v}F(u)}{v}+\left[ u^{4}-\gamma ^{2}u_{h}^{4}\right]
\frac{\dot{v}\zeta ^{\prime }(u,v)}{vu_{h}^{2}}\right\}   \notag \\
&&+T_0R^2\left( \pi T\right) ^{2}\frac{v^{2}}{\sqrt{%
1-v^{2}}}\left\{ \left[ \frac{\gamma ^{2}\ u^{4}-u_{h}^{4}}{u^{4}-u_{h}^{4}}%
\right] \frac{\ddot{v}\zeta (u,v)}{v}+\left[ u^{4}-\gamma ^{2}u_{h}^{4}%
\right] \frac{\ddot{v}\chi _{2}^{\prime }(u,v)}{vu_{h}^{2}}\right\}   \notag
\\
&&+T_0R^2\left( \pi T\right) ^{2}\frac{v^{2}}{\sqrt{%
1-v^{2}}}\left\{ \dot{v}^{2}\gamma ^{2}\left[ \frac{F^{2}\left( u\right) }{%
2f\left( u\right) }-\frac{\zeta ^{\prime 2}(u,v)u^{4}f\left( u\right) }{2}%
\right] \right\}   \notag \\
&&+T_0R^2\left( \pi T\right) ^{2}\frac{v^{2}}{\sqrt{%
1-v^{2}}}\left\{ \dot{v}^{2}\gamma ^{2}\left( F\left( u\right) +\frac{\zeta
^{\prime }(u,v)u^{4}f\left( u\right) }{u_{h}^{2}}\right) \left(
-u_{h}^{2}\zeta ^{\prime }(u,v)+\frac{F\left( u\right) }{f\left( u\right) }%
\right) \right\}   \notag \\
&&+T_0R^2\left( \pi T\right) ^{2}\frac{v^{2}}{\sqrt{%
1-v^{2}}}\left\{ \dot{v}^{2}\frac{F\left( u\right) \zeta ^{\prime
}(u,v)u^{4}f\left( u\right) }{v^{2}u_{h}^{2}}+\left[ u^{4}-\gamma
^{2}u_{h}^{4}\right] \frac{\dot{v}^{2}\chi _{1}^{\prime }(u,v)}{vu_{h}^{2}}+%
\frac{\dot{v}^{2}}{v}\frac{\partial \zeta (u,v)}{\partial v}\frac{\gamma
^{2}u^{4}-u_{h}^{4}}{u^{4}-u_{h}^{4}}\right\}\ .
\end{eqnarray}
The energy flow at $u=u_{s}$ does not depend on $\chi'_{1}\left( u\right)$
or $\chi'_{2}\left( u\right) $, and is found to be (in the large $\gamma$ limit where
the second order correction to $\dot{v}$ is subdominant)
\begin{equation}
\Phi(t,u_s)\underbrace{\simeq}_{\gamma \gg 1}T_0R^2\ u_h^2\gamma v^2
\left[1+\frac{1}{3\gamma ^{3/2}}\frac{u_h}{u_m-u_s}
+\left( \frac{\pi }{8}-\frac{1}{4}\log 2-\frac{7}{32}\right)
\gamma\left(\frac{u_h^2}{(u_m-u_s)^2}\right)
+{\cal O}\left(\frac{u_h^3}{(u_m-u_s)^3}\right)\right]\ ,
\label{eloss2nd}
\end{equation}
where we have used the results
\begin{equation}
\frac{\partial \zeta (u_{s},v)}{\partial v}=-v\gamma ^{3}\frac{1}{2u_{h}^{2}}%
\left( \frac{1}{4}+\frac{\pi }{4}-\frac{1}{2}\log 2\right) .
\end{equation}%

In order to rewrite this in terms of the heavy quark mass, and other gauge theory quantities,
one should first determine $M$ up to second order. From formula \eqref{E_Stat}, one finds
\begin{equation}
M=\frac{\sqrt{\lambda}}{2\pi}(u_m-u_s)\ \left(1-\frac{v^2 u_h^2}{u_m-u_s}F(u_s)\right)
\underbrace{\simeq}_{\gamma\gg1}\frac{\sqrt{\lambda}}{2\pi}(u_m-u_s)\
\left(1+\frac{v^2 u_h}{3\gamma^{3/2}(u_m-u_s)}\right)\ ,
\end{equation}
which is again consistent with $-Md\gamma/dt=\Phi(t,u_s).$ Then in formula
\eqref{eloss2nd}, when introducing
$(u_m-u_s)^{-1}=\sqrt{\lambda}/(2\pi M) \left[1+{\cal O}(u_h/(u_m-u_s))\right],$ one sees that what was first order in $u_h/(u_m-u_s)$ will introduce a second order correction in terms of $\sqrt{\lambda}T/(2M).$ However, in the large $\gamma$ limit, this contribution is subdominant.
Therefore one gets for the radiative energy loss
\begin{eqnarray}
\Phi(t,u_s)\underbrace{\simeq }_{\gamma \gg 1}\frac{\sqrt{\lambda }}{2\pi }\left(
\pi T\right) ^{2}\gamma v^{2}\left[ 1+\frac{1}{3\gamma ^{3/2}%
}\frac{\sqrt{\lambda }T}{2M}+\left( \frac{\pi }{8}-\frac{1}{4}\log 2-\frac{7%
}{32}\right) \gamma \left(\frac{\lambda T^2}{4M^2}\right)
+{\cal O}\left(\frac{\lambda^{3/2} T^3}{8M^3}\right)\right]\ .
\end{eqnarray}%
The second order correction is also positive ($\frac{\pi }{8}-\frac{1}{4}\log
2-\frac{7}{32}\simeq 0.00066$) but very tiny. This continues to enhance the
energy loss for heavy quark with lighter mass. Also, one can check that the
second order correction is always small as long as $u_{s}=\sqrt{\gamma }%
u_{h}\ll u_{m}$. It is also interesting to notice that the second
order correction is proportional to $\frac{\sqrt{\lambda }}{2\pi }a^{2}$
\cite{Mikhailov:2003er,Xiao:2008nr} when $a$ is identified as $\gamma^{3}\dot{v}$ in
relativity. One can interpret this term energy lost by radiation due to the deceleration.

\section{Thermalization time}

To finish, let us estimate the thermalization time of the heavy quark, and show that
our perturbative expansion can be trusted, provided $u_m$ is large enough and the heavy quark slows down in a quasistatic way. In order to show how our calculation improves the previous estimations, let us rewrite formula \eqref{vdot2} in the following way:
\begin{equation}
(\mu-\sqrt{\gamma})\frac{dv}{d\tilde{t}}=-v(1-v^2)
\left[1-\frac{\gamma}{(\mu-\sqrt{\gamma})^2}
\left(\frac14+\frac{\pi}8-\frac{\log 2}4\right)\right]\ ,
\label{vdoteq}\end{equation}
where $\tilde{t}=\pi T\ t$ and $\mu=u_m/u_h=2m/(\sqrt{\lambda}T)$ is expressed in terms
of the bare mass of the heavy quark. Note that $\mu\gg\sqrt{\gamma}$ in order for our perturbative expansion to be valid.

In Ref.~\cite{Herzog:2006gh,Gubser:2006bz}, the equation proposed for $\dot{v}$ is
$\mu\ dv/d\tilde{t}=-v(1\!-\!v^2)$ and the solution is 
\begin{equation}
v(t)=\sqrt{\frac{v_{0}^2}{v_{0}^2+\left(1-v_{0}^2\right)\exp\left(\frac{2\pi T t}{\mu}\right)}}\ ,\label{naive}\end{equation}
with $v_0\!=\!\sqrt{1\!-\!1/\gamma_0^2}$ the initial velocity. This naturally implies the
thermalization time scale to be $\frac{2m}{\sqrt{\lambda}\pi T^2}.$ However the equation proposed in Ref.~\cite{Herzog:2006gh,Gubser:2006bz} is different from our first order equation
$(\mu\!-\!\sqrt{\gamma})\ dv/d\tilde{t}=-v(1\!-\!v^2),$ which indicates that the thermalization time is rather given in terms of the thermal mass of the quark:
\begin{equation}
\tau=\frac{2M}{\sqrt{\lambda}\pi T^2}\ .
\label{thermtime}
\end{equation}
Strictly speaking $M$ slightly varies with $t,$ from $m-T\sqrt{\lambda\gamma_0}/(2\pi)$
to about $m-T\sqrt{\lambda}/(2\pi),$ which bounds the thermalization time. In any case it is
a smaller time than $\frac{2m}{\sqrt{\lambda}\pi T^2},$ and this leads to a faster thermalization. 

Let us point out that this difference comes from the fact that we had to pick a particular point on the string to evaluate the energy loss, while in the stationary problem solved in
Ref.~\cite{Herzog:2006gh,Gubser:2006bz}, it doesn't matter at what value of $u$ the energy flow is computed because the flow is constant when $\dot{v}=0.$ In our study, we actually
let the heavy quark slow down and the point on the string where the energy loss is evaluated becomes crucial. Our choice to evaluate the radiative energy loss at $u\!=\!u_s$ is motivated by the fact that the part of the string below $u_s$ is not causally connected to the part of the string above \cite{Dominguez:2008vd}, and therefore corresponds to emitted radiation on the gauge theory side.

\begin{figure}
\begin{center}
\includegraphics[width=12cm]{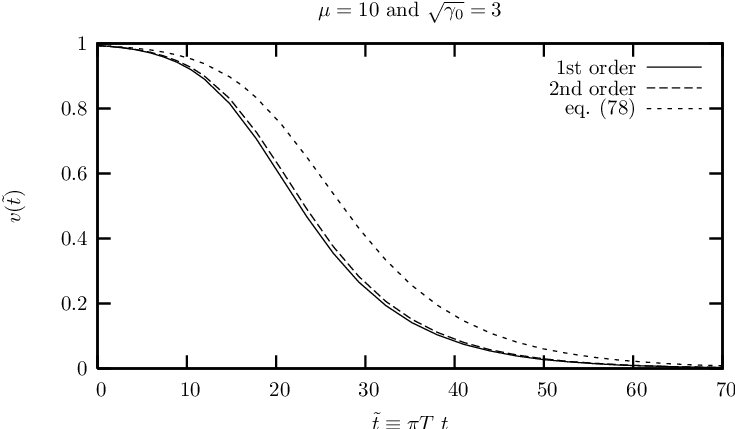}
\end{center}
\caption{Numerical solutions of equation \eqref{vdoteq} without the square bracket term (first order), with the square bracket term (second order), and the solution \eqref{naive} proposed in \cite{Herzog:2006gh,Gubser:2006bz}. Our results indicate a faster heavy quark thermalization.}
\label{acstring}
\end{figure}

In Fig.1, equation \eqref{vdoteq} is solved numerically at first order (without the square bracket term) and at second order. The comparison with the solution \eqref{naive} confirms that our calculations imply a faster thermalization: the time it takes to reduce the initial speed in half is about $2\mu/(\pi T),$ instead of $3\mu/(\pi T).$ We also see that the second order correction is quite small, although one should keep in mind that the dashed curve should be modified for small values of $\gamma:$ the second order correction in \eqref{vdoteq} is only valid in the
limit $\gamma\gg1.$


\section{Conclusion}

In this paper, we study the radiative energy loss of a slowly decelerating
heavy quark moving through a strongly coupled SYM plasma by using AdS/CFT
correspondence. The calculation is done in terms of perturbation in
$\sqrt{\lambda} T/M,$ where $\lambda$ is the strong coupling of the SYM theory,
$T$ is the temperature of the plasma and $M$ is the thermal mass of the heavy quark.
The shape of the string trailing behind the heavy quark in the fifth dimension
determines the medium-induced energy loss of the quark. At zeroth order, the heavy quark
propagates at a constant speed, and the trailing string \eqref{Ansatz_0thOrder}-\eqref{sol_F} is
that found in Ref.~\cite{Herzog:2006gh,Gubser:2006bz}. We determine the corrections to the shape of the string up to first order, formulae \eqref{Ansatz_1stOrder}, \eqref{sol_zeta} and 
\eqref{vdot_1stOrder}, and the radiative energy loss of the heavy quark up to second order
\eqref{eloss2nd}. In this case we only gave the large $\gamma$ result explicitely, but the exact expression was also found. The physical interpretation of each correction is provided. Especially, the second order correction
can be understood as energy loss due to the decelaration. Last but not least, these calculations allow us to evaluate the thermalization time of the heavy quark moving in the strongly-coupled SYM plasma. Our parametric estimate is given by \eqref{thermtime}.


\begin{acknowledgments}
We would like to thank Prof. A.H. Mueller for numerous discussions and
helpful comments. G.B. acknowledges helpful discussions with Edmond Iancu, Robi Peschanski
and Fran\c cois Gelis. G.B. would like to thank Prof. A.H. Mueller and other members of the Department of Physics of Columbia University for hospitality and support during the
early stage of this work. B.X. would like to thank Prof. Bo-Qiang Ma for hospitality
and support during his visit to the Physics Department of Peking University
when this work was finalized. C.M. is supported by the European Commission under
the FP6 program, contract No. MOIF-CT-2006-039860. B.X. is supported by the
US Department of Energy.
\end{acknowledgments}


\appendix

\section{Velocity of the quark $vs.$ $v(t)$}

Using our Ansatz \eqref{Ansatz}, the velocity of the heavy quark, which is the velocity of the string at the flavor brane, is
\begin{equation}
\dot{x}(t,u_m)=v(t)+\dot{v}(t)F(u_m)+\ddot{v}(t)\zeta (u_m,v)+\dot{v}^{2}(t)\partial_v\zeta (u_m,v)+ \dots\ .
\end{equation}
On the other hand, we have the expansions
\begin{eqnarray}
F(u_m)&=&-\frac{u_h^2}{3u_m^3} +{\cal O}\left(\frac{u_h^6}{u_m^7}\right)\ ,\\
\zeta(u_m,v)&=&-\frac{\gamma^2}{2 u_m^2}+{\cal O}\left(\frac{u_h}{u_m^3}\right)\ .
\end{eqnarray}
Therefore, the velocity of the quark reads
\begin{equation}
\dot{x}(t,u_m)=v(t)+{\cal O}\left(\frac{u_h^4}{u_m^4}\right)\ .
\end{equation}
The discrepancy between $v(t)$ and the velocity of the quark appears only at higher orders in the large mass expansion than the ones considered in this study.

\section{Checking energy conservation up to second order}
\label{ehigh}

According to Eq.~(\ref{edensity}) the energy of the system is
\begin{equation}
E=\int_{u_{s}}^{u_{m}}du\ e(t,u)=T_{0}R^{4}\int_{u_{s}}^{u_{m}}du\
\frac{1+u^{4}f(u)x^{\prime 2}}{\sqrt{-g}}\ .\label{defE1}
\end{equation}%
Inserting Eq.~(\ref{Ansatz}), one finds
\begin{equation}
\frac{1+u^{4}f(u)x^{\prime 2}}{\sqrt{-g/R^{4}}}=\gamma \left[ \left( 1+\frac{%
v^{2}u_{h}^{4}}{u^{4}-u_{h}^{4}}\right) +\dot{v}v\gamma ^{2}\left( 1+\frac{%
v^{2}u_{h}^{4}}{u^{4}-u_{h}^{4}}\right) \left( \frac{F\left( u\right) }{%
f\left( u\right) }-u_{h}^{2}\zeta ^{\prime }(u,\gamma )\right) +2\dot{v}%
vu_{h}^{2}\zeta ^{\prime }(u,v)+\mathcal{O}\left( \frac{\dot{v}^{2}}{%
u_{h}^{2}}\right) \right] \ ,  \label{energydensityfactor}
\end{equation}%
and thus, neglecting terms suppressed by powers of $u_{m}$, one gets
\begin{eqnarray}
E=T_0R^2\gamma \left\{u_{m}-u_{s}-v^{2}u_{h}^{2}F(u_{s})
-v\dot{v}\Big[-(\gamma^2-2)u_h^2\zeta(u_s,v)+\gamma^2\sigma(u_s)
+(\gamma^2-1)\frac{u_h^2}2F^2(u_s)\right.\nonumber\\\left.
-\gamma^2(\gamma^2-1)u_h^2\rho(u_s)\Big]
+\mathcal{O}\left( \frac{\dot{v}^2}{u_h^2}\right) \right\}\ .
\label{E2order}
\end{eqnarray}%
Using
\begin{eqnarray}
\frac{\delta \left. E\right\vert _{du_{s}}}{dt}=-e(t,u_s)\
\frac{u_s\gamma ^{2}v\dot{v}}{2}=
T_0 R^2\ v^2 u_h^2\left\{\frac{(\gamma^2+1)}{2\gamma}\frac{u_s}{u_m-u_s}\ 
\hspace{5cm}\right.\nonumber\\\left.
-\frac{v^2 u_s^3}{2\gamma^2(u_m-u_s)^2}\left[\frac{1+\gamma^4}{\gamma^2-1}F(u_s)
-\frac{\gamma(\gamma^2-1)}{4u_s}\right]
+{\cal O}\left(\frac{u_h^3}{(u_m-u_s)^3}\right)\right\}\ ,
\end{eqnarray}
and substracting $\Phi(t,u_s)$ (formula \eqref{eloss2nd}), one can check that the energy conservation equation is verified, meaning that one recovers the net energy loss $dE/dt$ from derivating \eqref{E2order}.



\end{document}